\documentclass[floatfix,twocolumn,showpacs,prb,amsmath,amssymb]{revtex4}

\usepackage{graphicx}
\usepackage{bm}

\begin{document}

\hyphenation{GaMnAs}
\hyphenation{GaMnP}
\hyphenation{GaMnN}
\hyphenation{GaMnSb}
\hyphenation{CdMnTe}
\hyphenation{CdMnSe}
\hyphenation{PbSnMnTe}

\title{
Ferromagnetic and random spin ordering
in diluted magnetic semiconductors
}

\author{A. Kaminski}
\author{V.M. Galitski}
\author{S. \surname{Das Sarma}}
\affiliation{
Condensed Matter Theory Center,
Department of Physics, University of Maryland, College Park, Maryland
20742-4111}

\begin{abstract}
In a diluted magnetic semiconductor system, the exchange interaction
between magnetic impurities has two independent components: direct
antiferromagnetic interaction and ferromagnetic interaction mediated
by charge carriers. Depending on the system parameters, the ground
state of the system may be ordered either ferromagnetically or
randomly. In this paper we use percolation theory to find the
ferromagnetic transition temperature and the location of the quantum
critical point separating the ferromagnetic phase and a valence bond
glass phase.
\end{abstract}

\pacs{75.50.Pp, 75.10.-b, 75.30.Hx}

\maketitle

\section{Introduction}\label{sec:intro}

In diluted magnetic semiconductors (DMS), e.g. Ga$_{1-x}$Mn$_x$As,
In$_{1-x}$Mn$_x$As etc, long-range ferromagnetic ordering of the
impurity (i.e. Mn) local moments is induced by the carriers (holes in
most cases) which are contributed by the dopant impurities serving the
dual role of acceptors and magnetic atoms. In principle, each Mn
impurity atom contributes one hole, but heavy compensation
intrinsically present in the system leads to a ``dilute'' carrier
system with $n_{\textrm{i}} \gg n_{\textrm{h}}$, where
$n_{\textrm{i}}$ and $n_{\textrm{h}}$ are the active local moment
density and carrier density respectively. Throughout this paper, we
refer to the carriers as ``holes'' and the magnetic local moments as
``impurities'' or ``Mn atoms'' without any loss of generality. The
origin of this heavy compensation is essentially unknown although As
antisite and Mn interstitial defects, both acting as electron
double-donors, are generally thought to be responsible. There is wide
consensus that the ferromagnetic ordering of the impurity moments at
low enough temperatures is caused by the local exchange interaction
between the impurity atoms and the holes. This exchange coupling
produces a long-range effective ferromagnetic interaction between the
impurity local moments, leading to the ferromagnetic state of the
system at temperatures $T < T_c$, where $T_c$ is the ferromagnetic
transition temperature. The precise mechanism for the carrier-mediated
magnetic coupling between the impurity atoms is still being discussed
and debated in the literature, particularly for the metallic DMS
systems where both free-carrier weak-coupling RKKY-Zener
interaction\cite{AbrikosovGorkov63,sds3} and strong-coupling
Zener-double-exchange interaction\cite{c3} have been invoked depending
on whether the holes reside in the semiconductor valence band or in
the Mn-induced impurity band respectively. The issue of whether the
carriers reside in the semiconductor valence band or in the impurity
band in the semiconductor band gap has been controversial.\cite{Timm}
The hole binding energy ($\sim 150$ meV above the valence band edge)
contributed by Mn in GaMnAs is fairly deep in the band-gap, and
therefore impurity band physics is certainly operational in the
strongly localized insulating DMS regime where the physical situation
must be of carriers strongly bound to randomly localized magnetic
impurities in the system. In our earlier
publications\cite{KaminskiDasSarma2002,add1,add2,DasSarmaHwangKaminski2002,GalitskiEtal2003}
we have developed a polaron percolation theory for DMS impurity
ferromagnetism in the strongly localized insulating regime by showing
that the localized holes produce magnetic clusters of bound magnetic
polarons with the polaron size increasing with decreasing temperature
until the magnetic clusters overlap through the whole sample leading
to the ferromagnetic temperature-induced polaron percolation
transition (more details on this scenario are provided in
Sec.~\ref{sec:perc}).  Such a magnetic polaron transition scenario has
later been explicitly verified in direct numerical simulations of the
DMS Hamiltonian.\cite{Dagotto} In the current work we generalize our
polaron percolation picture by explicitly including in the theory the
direct (i.e. not carrier-mediated) antiferromagnetic exchange
interaction between the impurity atoms which play an important role in
II-VI materials, where the direct antiferromagnetic exchange
interaction is strong and at larger impurity concentrations in III-V
materials, when the magnetic impurities are more likely to be at short
distances from each other.

The DMS materials are interesting systems from the fundamental
perspective of frustration physics because there are independent
competing ferromagnetic and antiferromagnetic contributions to the
magnetic Hamiltonian arising from totally distinct physical processes:
the ferromagnetic contribution is carrier-mediated and effectively
long-range whereas the antiferromagnetic contribution is extremely
short-ranged direct Mn-Mn exchange. The disorder inherent in random
locations of the impurity atoms (i.e. Mn in Ga$_{1-x}$Mn$_x$As)
coupled with the competing long-ranged (carrier-mediated)
ferromagnetic and short-ranged direct antiferromagnetic interaction
among the impurity atoms leads to interesting frustration physics in
the system. This frustration will eventually lead to a
non-ferromagnetic ground state (possibly, a spin glass state) in the
system at high enough impurity concentration and we study this DMS
frustration physics within our polaron percolation model. We believe
that this random competition between ferromagnetic and
antiferromagnetic coupling leads to the disappearance of DMS
ferromagnetism at high impurity concentration and is responsible for
the generic absence of ferromagnetism (with a few exceptions with very
low $T_c$ values) in II-VI DMS materials. As described in the rest of
this paper, we find that $T_c$ shows a maximum as a function of the
magnetic impurity density $n_{\textrm{i}}$ due to this random
competition, and for large enough impurity density the system has a
non-ferromagnetic ground state. We would like to mention in this
context Ref.~\onlinecite{SGexp}, in which a spin glass state was
observed in an insualting $\left({\rm Ga},{\rm Mn}\right){\rm N}$
alloy (which is a III-V dilute magnetic semiconductor). It is possible
that the spin glass freezing is due to the random antiferromagnetic
coupling between magnetic impurities, which becomes dominant at high
impurity concentration.

The outline of the paper is as follows: The model is introduced in
Sec.~\ref{sec:model}. In Sec.~\ref{sec:perc} we describe the
percolation theory approach we use to deal with the problem at hand.
In Sec.~\ref{sec:Tcsupp}, we investigate suppression of the
ferromagnetic transition temperature by direct antiferromagnetic
interaction between the Mn impurities. The suppression is due to the
fact that the coupling of spins to charge carriers and the strength of
the interaction between spins strongly depend on the concentration of
the magnetic atoms.  We show that the approach used in
Sec.~\ref{sec:Tcsupp} fails at very large concentrations of Mn atoms.
This case of large Mn concentrations is considered in
Sec.~\ref{sec:zero-temp}, where a different technique is used. The
results of the paper are summarized in Sec.~\ref{sec:concl}.

\section{The model}\label{sec:model}
The magnetic Hamiltonian of the system has the form
\begin{equation}
\label{H}
\hat{H}=\sum_{jk} J(r_{jk}) \textbf{S}_j \textbf{s}_k
+\sum_{j_1 j_2} J^{\textrm{AF}}(r_{j_1j_2}) \textbf{S}_{j_1}
\textbf{S}_{j_2}\;, 
\end{equation}
where indices $j$ and $k$ run over magnetic impurities and localized
holes respectively, $\textbf{S}_j$ and $\textbf{s}_k$ denote spins of
magnetic impurities and localized holes, $J(r_{jk})$ is the constant
of impurity-hole exchange interaction, with its dependence on the
distance $r_{jk}$ between the hole localization center and the
magnetic impurities given by
\begin{equation}
\label{J}
J(r) = J_0 \exp\left(-\frac{r}{L}\right)\;,
\end{equation}
with $L$ being the hole localization radius. The precise origin for
the hole localization is irrelevant for our theory -- it could be
Coulombic binding to the impurity acceptor or disorder-induced
Anderson localization or any other relevant localization mechanism.
The sign of the impurity-hole exchange interaction does not matter for
the physical properties of the impurity subsystem, since the effective
exchange interaction between impurities induced by the first term of
Hamiltonian~(\ref{H}) is always ferromagnetic (see, for example,
Ref.~\onlinecite{KaminskiDasSarma2002}). We choose $J_0>0$ (that is
antiferromagnetic) in our consideration without any loss of generality
simply because that happens to be the case for the exchange coupling
between a hole and a Mn moment in GaMnAs.

The direct antiferromagnetic
interaction between magnetic impurities also decays exponentially with
the distance,
\begin{equation}
\label{JAF}
J^{\textrm{AF}}(r) = J^{\textrm{AF}}_0 
\exp\left(-\frac{r}{a}\right)\;,
\end{equation}
and the corresponding decay length $a$ is much smaller than that of
the impurity-hole interaction, $a\ll L$, since the former is
determined by the (small) decay length of the electron wave function of an
inner shell of the magnetic impurity, while the latter is determined
by the (larger) decay length of the wave function of an impurity-band
carrier localized at some defect in the crystal structure.

Parameters of Hamiltonian~(\ref{H}), $J_0$, $J_0^{\textrm{AF}}$, $a$,
and $L$, together with impurity and hole concentrations
$n_{\textrm{i}}$ and $n_{\textrm{h}}$ make the full set of parameters
determining the properties of the system. In a realistic diluted
magnetic semiconductor system, these parameters obey the following
relations:
\begin{subequations}
\label{relations}
\begin{equation}
\label{relations-a}
\frac{1}{n^{1/3}_{\textrm{h}}} \gg \frac{1}{n^{1/3}_{\textrm{i}}} \gg
a\;,\quad 
L \gg a\;.
\end{equation}
The fact that the charge carriers in the system under consideration
are localized means that
\begin{equation}
\label{relations-b}
\frac{1}{n^{1/3}_{\textrm{h}}}  \gg L \;.
\end{equation}
\end{subequations}
The relations between the pairs of parameters
\begin{equation}
\label{relations-c}
\frac{1}{n^{1/3}_{\textrm{i}}}\ \textrm{and}\ L\;, \quad
n^{1/3}_{\textrm{h}} L \ \textrm{and}\ 
n^{1/3}_{\textrm{i}} a
\end{equation}
are not predetermined in our consideration. 

In this work, we will treat both impurity and hole spin dynamics
classically. For impurity spins, it is justified by their relatively
large value $S=5/2$ in real experimental systems, such as GaMnAs. The
hole spins, even though not so large ($s=3/2$ in GaMnAs), are strongly
coupled to the magnetic impurities closest to the hole localization
centers, which makes them ``massive'' (that is effectively classical)
as far as their interaction with more distant impurities is
concerned.\cite{KaminskiDasSarma2002} In any case, the quantum
properties are unlikely to be of qualitative importance in determining
$T_c$. 

For analytic description of the system described by
Eqs.~(\ref{H})--(\ref{relations}), it turns out to be important
whether relation
\begin{equation}
\label{relations-d}
L 
\frac{1}{\left(L n^{1/3}_{\textrm{h}}\right)^{1/3}} \gg 
\frac{1}{n^{1/3}_{\textrm{i}}}
\end{equation}
is obeyed or not. The former case is more interesting physically, so
it will be considered in this paper. The latter case requires
a completely different formalism to treat but yields qualitatively
similar results. In what follows we confine ourself to the former
case, assuming that the inequality~(\ref{relations-d}) holds.

\section{Percolation theory approach}\label{sec:perc}
\subsection{Basic principles}\label{sec:perc-basic}

To understand the physics of ferromagnetic transition in a diluted
magnetic semiconductor system, it is easier first to consider the
system at temperatures well above the ferromagnetic transition
temperature, and then follow the evolution of the system as the
temperature is lowered. So at high enough temperatures ($T\gg J_0$)
spins of holes and magnetic impurities are not correlated. As the
temperature goes down, magnetic impurities close enough to hole
localization centers tend to align their spins in the direction
opposite to that of the hole they are close to.  The resulting complex
of a hole and magnetic impurities polarized by it is called a ``bound
magnetic polaron.''\cite{Durst} The radius of a bound magnetic polaron
grows as $T\to 0$.  When two or more bound magnetic polarons overlap,
their spins must have the same direction. These overlapping polarons
make polaron clusters; the ferromagnetic transition occurs when a
macroscopic ``infinite cluster'' appears. In our earlier
works\cite{KaminskiDasSarma2002,GalitskiEtal2003} we considered this
ferromagnetic transition in the absence of antiferromagnetic
interaction between magnetic impurities ($J_0^{\textrm{AF}}=0$). While
this antiferromagnetic interaction certainly suppresses ferromagnetic
transition in the system (to what extent, we will see later), the
basic mechanics of the transition remains the same: bound magnetic
polarons grow, merge, make polaron clusters, and finally formation of
an infinite cluster signals ferromagnetic transition. We note that the
DMS magnetic polaron percolation transition is not the usual
concentration-driven $T=0$ percolation transition, but a more subtle
temperature-driven transition involving the coalescence of magnetic
polaron clusters.

Thus a bound magnetic polaron is the basic unit of the mechanism of
ferromagnetic transition in diluted magnetic semiconductors with
localized charge carriers. We demonstrated in our earlier
paper\cite{GalitskiEtal2003} that in the absence of antiferromagnetic
interaction between the impurities they can be integrated out provided
Eq.~(\ref{relations-d}) holds, and the effective energy of the
interaction between two polarons with distance $r$ between their
centers can be obtained:
\begin{eqnarray}
\label{Jeff}
\left.{\cal E}_{\rm eff}\left( r, \cos\theta, T
 \right)\right|_{J_0^{\textrm{AF}}=0} 
 = &-&  
{2 \pi \over 3} 
\left[  L^2 r n_{\textrm{i}}\ln \left( {T \over J_0}
  e^{r \over 2 L }\right) \right]
 \nonumber \\ 
&&\times  {J_0^2 \over T}
  \exp{\left( -{r \over L} \right)} \cos{\theta}
\end{eqnarray}
where $\theta$ is the angle between the spins of two polarons. One can
see that the interaction between the polarons decays exponentially
with distance between them. This reduces the problem of ferromagnetic
transition in a system of bound magnetic polarons to the analogous
problem of randomly placed spins with ferromagnetic exchange interaction
between them decaying exponentially. 

The latter problem was considered by Korenblit \emph{et
  al.}\cite{KorenblitEtAl73} in the context of ferromagnetic phase
transitions in diluted magnetic alloys. They proposed mapping of the
physical problem at hand to the mathematical problem of percolation
transition of same-radius randomly-placed spheres. Such a mapping is
possible because of two factors: the exponentially fast decay of
interaction between the spins and the characteristic distance between
spins being much larger than the decay length of the interaction [see
Eq.~(\ref{relations-b})]. Under these conditions, the distribution of
the nearest-neighbor couplings is wide. This, in turn, means, that an
any given temperature $T$ only a small fraction of
nearest-neighbor-coupling strengths is of the order of $T$, and all
the others are either much stronger or much weaker. The basic idea of
the approach of Korenblit \emph{et al.}\cite{KorenblitEtAl73} is that
we can postulate all spins with coupling strengths between them
stronger that $T$ to be locked in the same direction, and completely
neglect all the other couplings, which, in general, are exponentially
weaker. The justification for this approach is that there are very few
couplings of the order of temperature, and their influence on the
general picture is negligible; the validity of this justification will
be discussed in the next subsection.

The formal mapping to the problem of percolating spheres is done in
the following way:\cite{KorenblitEtAl73} all spins are replaced with
spheres of the same radius $R$, which is half the distance between
two spins whose coupling exactly equals temperature. When two
spheres (\emph{i.e} spins) are closer than $2R$, they overlap (and
the coupling between the corresponding spins is larger than $T$).
The spheres that overlap make clusters (which are magnetic clusters
discussed above), and the ferromagnetic transition temperature is
reached when the sphere radius (which grows as the temperature goes
down) becomes large enough for the infinite cluster to appear. This
problem of overlapping spheres has only one parameter, $nR^3$, where
$n$ is the sphere concentration, and can be easily solved
numerically. Mapping the results back to the original physical problem
of interacting spins, we get expressions for the ferromagnetic
transition temperature $T_c$ and other physical characteristics such
as magnetization, susceptibility
etc.\cite{KorenblitShender78,DasSarmaHwangKaminski2002} 
In the case of the system described by Hamiltonian (\ref{H}) with 
$J^{\textrm{AF}}_0=0$ the transition temperature is given
by\cite{KaminskiDasSarma2002,GalitskiEtal2003}
\begin{equation}
\label{Tc}
T_c\approx sS|J_0|  \left(L n_h^{1/3}\right)\! \sqrt{n_i/n_h}\
\exp\left(-\frac{0.86}{L n_h^{1/3}}\right)
\end{equation}

We note that the solution of the problem involves two distinct
mappings: first, the mapping of the DMS bound polaron Hamiltonian of
holes and impurities to the problem of random spins with ferromagnetic
exchange, and then the random spin problem to the mathematical
percolation problem of random spheres. We also note that there is no
characteristic temperature other than $T_c$ in this scenario.

\subsection{On the applicability of the percolation picture}
\label{sec:applic}

In this section we discuss the limits of applicability of the
percolation theory to the ferromagnetic transition in strongly
disordered systems. Despite being an extremely useful tool in dealing
with strongly disordered systems (sometimes, the only tool
applicable), the percolation theory, as applied to temperature-driven
phase transitions, has its drawbacks. Probably, the most important
drawback, which is the center of this subsection's discussion, is the
inability of the percolation theory to account for thermal
fluctuations. We would like to emphasize that this drawback is not
specific to the percolation of bound magnetic polarons, which we
consider in this and our earlier
papers.\cite{KaminskiDasSarma2002,DasSarmaHwangKaminski2002,GalitskiEtal2003}
In fact, any treatment of the temperature-driven ferromagnetic phase
transition in a strongly disordered using the percolation theory,
including the paper by Korenblit \emph{et al.},\cite{KorenblitEtAl73}
in which this treatment was first proposed, inevitably neglects thermal
fluctuations. In this section we argue that despite this problem, the
ferromagnetic transition temperature, as predicted by the percolation
theory, differs from the real transition temperature only by a numerical
factor of the order of unity.

We argue that the percolation picture of the temperature-driven
ferromagnetic phase transition in a strongly disordered system of
spins is qualitatively correct. When the temperature goes down,
ferromagnetic correlation is inevitably established first between the
most-strongly-coupled spins, then spreading across weaker couplings.
In such a system above the ferromagnetic transition temperature, one
has ferromagnetically ordered regions (ferromagnetic clusters) without
strong correlation between the magnetic moments of different clusters.
As the temperature goes down, more couplings become saturated, so the
new clusters appear and the existing ones grow and merge. Clearly at
some low enough temperature this growth and merging of clusters will
result in appearance of a correlated region which spans the whole
sample (infinite cluster). This scenario is referred to, for example,
in the context of the Griffiths phase physics, which was introduced
before the first application of the percolation theory to the problem
of ferromagnetic phase transitions. It seems to be universally
accepted as far as the qualitative description of the physics is
concerned. We have recently discussed the relevance of Griffiths phase
to DMS systems in the context of the magnetic percolation transition
in Ref.~\onlinecite{GalitskiEtal2003}.

However, it is hard to build a rigorous controllable approach based on
the above scenario. A reasonable approximation\cite{KorenblitEtAl73}
described in Sec.~\ref{sec:perc-basic} is to assume that at any given
temperature $T$ two spins connected by some exchange coupling $J$ are
completely uncorrelated if $J<T$, and locked in the same direction is
$J>T$. Clearly, thermal fluctuations are completely left out of this
approach. Despite this fact, such an approach should be adequate far
from the transition point, when the clusters formed by the interacting
spins have finite size. Any problems arising in estimating $T_c$ can
be taken care of by introducing numeric corrections of the order
of unity to the physical quantities calculated. The infinite cluster
is, however, a different matter. By definition, spins making the
infinite cluster must all be correlated across arbitrarily large
distances. Small but finite deviations from the strict alignment may
accumulate over large distances resulting in a complete loss of
long-range coherence. Still one should agree that the exponent in
Eq.~(\ref{Tc}) is predicted by the percolation theory correctly.
Indeed, there can be no ferromagnetic order at $T\gg T_c$ with $T_c$
given by Eq.~(\ref{Tc}), because no connected network of couplings
larger than such $T$ would exist. The connected network of such couplings
does appear at $T\sim T_c$. The transition temperature, however, is
determined not only by the strengths of the couplings making the
connected network, but also by the topology of the network itself. The
latter, however, is characterized only by power (not exponential)
dependence on the system parameters, as it follows from the scaling
assumptions of the percolation theory.\cite{StaufferBook} Therefore
the exponent in Eq.~(\ref{Tc}) is the only exponent which may enter
the expression for the true $T_c$.

Now we are going to argue that not only the exponent, but also the
parametric prefactor at the exponent for the true $T_c$ is correctly
predicted by the percolation theory. To begin with, let us note that
the relation between the parameter $p\equiv n^{1/3}_{\textrm{i}} r(T)$
of the percolation theory problem and the temperature reads
\begin{equation}
\label{p}
p=n^{1/3}_{\textrm{h}} L \log\left( \frac{J_0}{T}\right)\;,
\end{equation}
with $n_{\textrm{h}}^{1/3} L \ll 1$ [see Eq.~(\ref{relations-b})], and the
critical value of $p$ being $p_c\approx 0.86$. In the vicinity of the
transition point, $p\approx p_c$, we have
\begin{equation}
\label{ppc}
\frac{|p-p_c|}{p_c}\approx \frac{n^{1/3}_{\textrm{h}} L}{0.86}
\frac{|T-T_c|}{T_c} \ll \frac{|T-T_c|}{T_c}\;.
\end{equation}
Therefore, at $|T-T_c|/T_c\sim 1$, that is away from the thermodynamic
transition, we still have $|p-p_c|/p_c \sim n^{1/3}_{\textrm{h}} L \ll
1$, so the system is still near the percolation threshold as far as
the cluster topology and the scaling relations of the percolation
theory are concerned.

\begin{figure}
\includegraphics[width=3in]{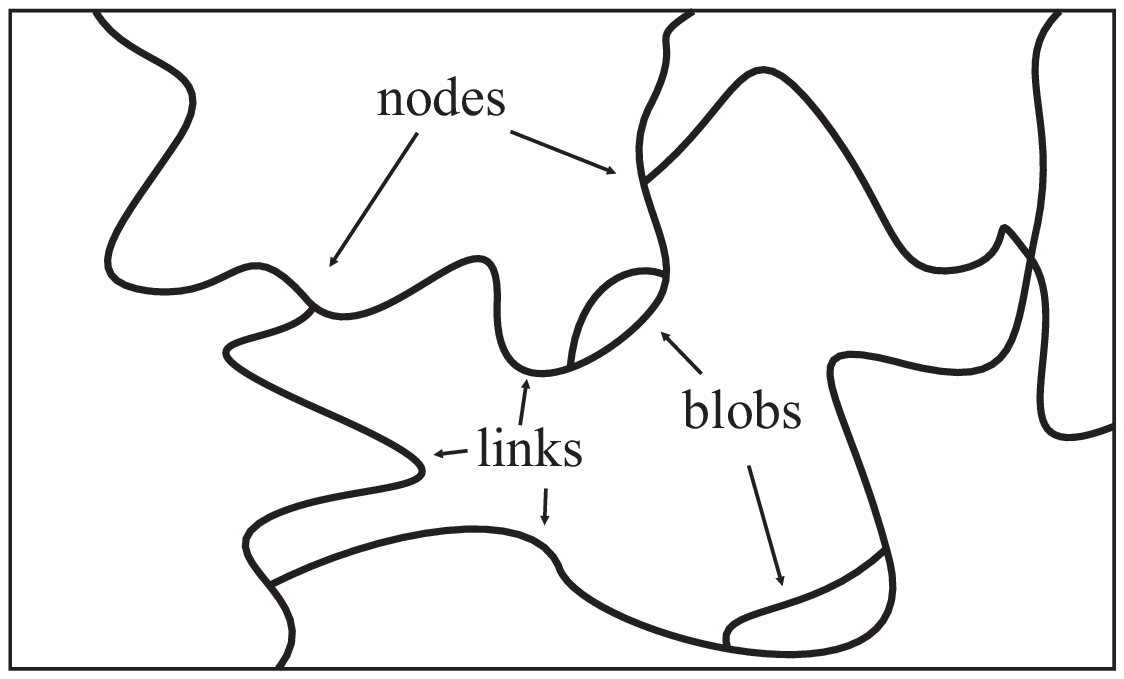}
\caption{\label{fig:top} 
Structure of the infinite cluster (after Refs.~[\protect\onlinecite{StaufferBook}] and
[\protect\onlinecite{EfrosShklovskiiBook}]).
 }
\end{figure}

Now let us consider the system of spins at some temperature $T = A
T_c$, where $A<1$ is a constant of the order of unity. Let us consider
the spins making up the infinite cluster (in terms of the percolation
theory) and argue that this infinite cluster must be in the
ferromagnetic state at some value of $A$, which is \emph{not}
parametrically smaller than unity. As the basis for our arguments, we
are going to use the ``links, nodes and blobs''
picture\cite{StaufferBook,EfrosShklovskiiBook} describing the topology of the infinite
cluster in the vicinity of the percolation transition, see
Fig.~\ref{fig:top}. According to this picture, the infinite cluster is
a network of ``links'' connecting ``nodes'', with some occasional
``blobs'' embedded into the links. The link length is proportional to
$|p-p_c|$ to some power of the order of unity. One may argue that the
transition temperature of such a network is of the order of
$J_{\textrm{link}}$, where $J_{\textrm{link}}$ is the effective
exchange interaction between the nodes as it is transmitted by the
link. The distribution of the couplings along the link is wide, the
weakest coupling is exponentially small as compared to the average
one, and the length of the link is large only as a power of the same
parameter that enters the exponent. In such a system, it is the
weakest coupling which gives the value of $J_{\textrm{link}}$, and
therefore the critical temperature must be of the order of the weakest
coupling in a link, which is of the order of $T_c$ as given by
Eq.~(\ref{Tc}).  Therefore, the ``real'' critical temperature may not
be parametrically smaller that the one predicted by the percolation
theory.

Now let us argue that the ``real'' critical temperature should not be
parametrically larger than the percolation theory prediction. In
principle, there is only one link of the strength of
$J_{\textrm{link}}$ between two nodes, so if it was the only link at
all, the ferromagnetic temperature would have to be of the order of
$J_{\textrm{link}}$. Even though other links connecting the nodes are
weaker, they may significantly increase the critical temperature
provided there is a sufficient number of them. We, however, argue that
the number of weaker links is \emph{not} large enough to compensate
for their weakness. Indeed, the number of weaker links connecting two
nodes may be as large as at most some power of the scaling parameter
$|p-p_c|/p_c$, which at $T\sim T_c$ is of the order of
$n^{1/3}_{\textrm{h}} L $, see Eq.~(\ref{ppc}). The link strength, on
the other hand, is widely distributed due to the exponential decay of
the interaction and the probability of parametrically many links
having their strength of the order of $J_{\textrm{link}}$ is
exponentially small, so the connection between two nodes is dominated
by the strongest link connecting them, whose strength is
$J_{\textrm{link}}$, and the critical temperature may not be
parametrically higher than $J_{\textrm{link}}$.

This reasoning leads us to the conclusion that while the ferromagnetic
transition in the system under consideration is not, rigorously
speaking, a percolation transition, the result for $T_c$ given by the
percolation theory may differ from the real transition temperature
only by a factor of the order of unity. We do not make any claims that
our reasoning is a rigorous proof of our statement, but we do state
that it is coherent with the established notions of the percolation
theory\cite{StaufferBook} while the contrary statement would conflict
with these notions. The final answer may be given only by a numerical
simulation, and we point out that the Monte-Carlo
results\cite{WanBhatt00} on the ferromagnetic transition temperature
in diluted magnetic semiconductors with localized charge carriers is
in excellent agreement with our theory\cite{KaminskiDasSarma2002},
which is an indication that the reasoning of this section is correct.

\section{Suppression of $T_c$}\label{sec:Tcsupp}

Because the decay length $a$ of the inter-impurity antiferromagnetic
interaction is much less than the characteristic distance
$n^{1/3}_{\textrm{i}}$ between them, see Eq.~(\ref{relations-a}), the
distribution of the coupling strength (\ref{JAF}) between neighboring
impurities is wide. Assuming the positions of the impurities to be
random and uncorrelated, one can easily calculate the probability
$P(J)$ that antiferromagnetic coupling
$J^{\textrm{AF}}$ between an impurity and its nearest neighbor is
smaller than $J$:
\begin{equation}
P(J)=\exp\left[-\frac43\pi n_{\textrm{i}} a^3
\left(\ln\frac{J^{\textrm{AF}}_0}{J}\right)^3 \right]\;.
\label{PJ}
\end{equation}

It has been
demonstrated\cite{WolffEtAl96,KaminskiDasSarma2002,GalitskiEtal2003}
that ferromagnetic interaction between two bound magnetic polarons
(separated by the distance $r$) occurs mostly due to weakly-polarized
lens-shaped region between the polaron centers, with ``thickness'' 
$L\ln [ (T / J_0) \exp (r / 2 L)]$
and ``radius'' $\sqrt{Lr}$, see
Ref.~\onlinecite{KaminskiDasSarma2002,GalitskiEtal2003}. In the
expression (\ref{Jeff}) for the 
effective energy of the polaron interaction with
$J_0^{\textrm{AF}}=0$, the term in square brackets is essentially the
number of magnetic impurities in this lens-shaped region, whose (weak)
polarization by one polaron affects the other polaron and vice versa.
Non-zero antiferromagnetic interaction suppresses this interaction in
the following way: if the distance $R$ between two impurities in the
lens-shaped region is small enough, so that $J^{\textrm{AF}}(R) > T$,
the spins of these two impurities are locked in the opposite
directions, which means that they are not affected by the polarons and
drop out of the group of impurities transferring interaction between
the two polarons. The probability that an impurity is \emph{not}
excluded from this interaction transfer equals $P(T)$ with function
$P$ given by Eq.~(\ref{PJ}).  Therefore, the effective energy of
interaction of two bound magnetic polarons in the presence of (direct)
antiferromagnetic interaction between magnetic impurities is given by
Eq.~(\ref{Jeff}) with an extra factor of $P(T)$ in the right-hand
side, which accounts for the reduction of the number of impurities
mediating interaction between the two polarons [factor in brackets in
Eq.~(\ref{Jeff})].

Since the concentration of holes is relatively low, $n_{\textrm{h}}L^3
\ll 1 $, the relative variations of ${\cal E}_{\rm eff}\left( r,
\cos\theta,T \right)$ for different pairs of neighboring polarons are
very large. It enables us to use percolation theory to establish the
temperature of ferromagnetic transition in the system, as described in
Sec.~\ref{sec:perc}. According to the percolation
theory,\cite{KorenblitEtAl73,KaminskiDasSarma2002} the ferromagnetic
transition temperature $T_c$ is given by the condition
\begin{equation}
\label{Tccondition}
{\cal E}_{\rm eff}\left( 0.86 n_{\textrm{h}}^{\frac{1}{3}},
\cos\theta, T_c  \right) = T_c\;.
\end{equation}
For not very large $J_0^{\textrm{AF}}$, Eq.~(\ref{Tccondition}) can be
solved for $T_c$ self-consistently by iterations, which yields:
\begin{eqnarray}
T_c&\sim& sS|J_0|  \left(L n_h^{1/3}\right)\! \sqrt{n_i/n_h}\
\exp\left(-\frac{0.86}{L n_h^{1/3}}\right)\nonumber\\
&\times& \exp\left(-\frac{4\pi}{3}\,0.86^3 
\frac{a^3n_{\textrm{i}}}{L^3n_h}\right)
\;.
\label{Tciter}
\end{eqnarray}
This expression for $T_c$ becomes invalid when the antiferromagnetic
interaction between the impurities becomes so strong that most of them
are locked that is when $a^3n_{\textrm{i}} \gg L^3n_h$. The next
section deals with this case.

\section{Quantum phase transition}\label{sec:zero-temp} 

At some critical concentration of magnetic impurities, the Curie
temperature is suppressed to zero by the direct antiferromagnetic
interaction. This is a quantum phase transition point separating a
ferromagnetic state and a disordered state in the system of magnetic
impurities (with the density of magnetic impurities being the control
parameter).

Let us briefly and qualitatively discuss the possible structure of the
disordered state. It is reasonable to start with a model in which
there are only randomly placed magnetic impurities with
antiferromagnetic interaction (\ref{JAF}) between them, and no holes.
The classical ground state in the continuum model is not frustrated as
the interaction between impurities is a well-defined (exponential)
function of the distance and the latter is a random variable.  In the
continuum model, the probability of finding, for example, a cluster of
three impurities located in the corners of an equilateral triangle is
exactly zero. One can say that the strong Possionian disorder in the
continuum model lifts the degeneracy of the classical grounds state.

At low temperatures, the quantum nature of the spins of magnetic
impurities should reveal itself. To get an insight into the ground
state of the system one can use the idea of the real-space
renormalization group technique frequently used in the studies of low
dimensional antiferromagnetic systems.\cite{RSRG} In this method, one
systematically integrate out the strongest bonds gradually reducing
the energy scale at each step of the decimation procedure. The
strongest bonds are considered frozen into a singlet state.  The
locked impurity pairs are then removed from the system; however, the
couplings between other spins get renormalized due to virtual triplet
excitations across the strong bonds. The main assumption of the
real-space renormalization group technique is that physical properties
are determined by an effective low-energy distribution of couplings
(which in some cases has a universal form essentially independent of
the initial distribution).  The averaged characteristics of the spin
system (such as correlation functions) are determined by very rare
events, corresponding to a fraction of spins locked into a singlet
being separated by a very large distance.  These rare bonds dominate
long-range correlation functions. It is important to emphasize that
typical properties of the system are drastically different from the
averaged properties and essentially determined by the majority of
spins locked into a singlet with their nearest neighbours (one also
should keep in mind that Griffiths effects are much less pronounced in
higher dimensions). Therefore, the quantum ground state of our system
can be represented as a set of randomly oriented spin pairs (valence
bonds) locked into singlets. Such a state  can be called a {\em valence bond
  glass}.  Using this term, we can conclude that by increasing the
density of magnetic impurities in a DMS system, one may induce a phase
transition from a ferromagnetic (low impurity concentration) into a
valence bond glass state, in which the {\em typical} correlation
function decays, at the lengths of the order of characteristic
separation $n_{\textrm{i}}^{-1/3}$ between the impurities. At larger
scales, the local impurity spin orientation can be considered as a
random variable. 

The actual quantitative details for the quantum phase transition to
the valence bond glass phase are beyond the scope of the current work,
where we consider the (classical) competition between random
ferromagnetic and antiferromagnetic couplings induced by disorder at
high magnetic impurity concentrations. It is clear, however, that the
$T_c$-suppression given in Eq.~(\ref{Tciter}) leads to a glassy magnetic
state for large values of $n_i$, and quantum effects are obviously
important in the elucidation of this glassy state.

\subsection{Bound magnetic polaron in a random medium of magnetic impurities}

Now let us consider one hole localized among magnetic impurities,
still at zero temperature. Closer to the hole localization center, its
interaction with the impurities is stronger than the antiferromagnetic
interaction between the impurities, so all impurity spins in the
vicinity of the hole localization center are aligned in the same
direction. Away from the hole localization center, its interaction
with the impurities becomes weaker, and at some distance
$R_{\textrm{p}}$ the antiferromagnetic interaction between impurities
prevails over their interaction with the hole. To describe this
behavior quantitatively, we introduce $n_{\textrm{f}}(r)$, which is
the concentration of the impurities aligned by their interaction with
the hole spin rather than by their antiferromagnetic interaction with
neighboring impurities. $n_{\textrm{f}}(r)$ equals the product of the
impurity concentration $n_{\textrm{i}}$ and the probability $P[J(r)]$
that the coupling of and impurity to its nearest neighbor is weaker
than its coupling $J(r)$ to the hole. Using Eqs.~(\ref{J}) and
(\ref{PJ}), we arrive at
\begin{equation}
\label{nf}
n_{\textrm{f}}(r)=n_{\textrm{i}}\exp\left[-\frac43\pi a^3
n_{\textrm{i}} \left(\frac{r}{L}\right)^3\right]\;.
\end{equation}
The characteristic length $R_{\textrm{p}}$, at which
$n_{\textrm{f}}(r)$ decays is given by
\begin{equation}
\label{Rp}
R_{\textrm{p}} = \frac{L}{n_{\textrm{i}}^{1/3}a}\;,
\end{equation}
and will be called the ``zero-temperature radius of a bound magnetic
polaron.'' 

The magnetic impurities surrounding a bound magnetic polaron have some
preferred spin directions determined by their antiferromagnetic
interaction with their neighbors. A bound magnetic polaron ``feels''
this as an effective magnetic field $\textbf{B}_{\textrm{rand}}$,
whose amplitude and direction are random and are determined by the
(random) configuration of impurities around it. Now we will estimate
the characteristic magnitude of this field. There are two random
contributions to this field. One, $\textbf{B}^{(1)}_{\textrm{rand}}$,
comes from the interaction of the hole with the magnetic impurities
which are not polarized by it. The other contribution,
$\textbf{B}^{(2)}_{\textrm{rand}}$, is due to the interaction of the
impurities polarized by the hole with the unpolarized ones. The energy
coming from interaction of an impurity spin polarized by the hole with
other polarized impurities is also random, but it does not depend on
the hole's spin orientation, so it should not be included into the
effective magnetic field which represents the action of the random
impurity medium on the hole spin.
 
Rigorous analytic evaluation of these two contributions is hardly
possible. One could naively expect that they can be expressed in terms
of integrals of $n_{\textrm{f}}(r)$, $n_{\textrm{i}} -
n_{\textrm{f}}(r)$, and $J(r)$ over volume. However, while
$n_{\textrm{f}}(r)$ is indeed the concentration of impurities whose
spin direction is determined by the hole spin, $n_{\textrm{i}} -
n_{\textrm{f}}(r)$ is \emph{not} the concentration of randomly
oriented impurities that determine the magnitude and direction of
$\textbf{B}_{\textrm{rand}}$. The reason is that in addition to two
above mentioned categories of impurities -- the ones polarized by the
hole and the ones belonging to the random medium which extends away
from the holes -- there is a third category. An example of impurities
of this third category is given by two impurities at the distance $r_*
< R_{\textrm{p}}$ from the hole localization center with the distance
between them small enough for the antiferromagnetic interaction
between them to be stronger than $J(r_*)$. One can easily see that
these two impurities do not belong to those making
$n_{\textrm{f}}(r)$, but since this pair is able to rotate as a whole
it will orient itself in the most energetically favorable position
with respect to the spin of the hole, it will not contribute to
$\textbf{B}_{\textrm{rand}}$. We are not aware of any analytic way to
separate these impurities from the random continuum, and therefore we
must limit ourselves to estimates of $\textbf{B}_{\textrm{rand}}$
instead of a rigorous evaluation.

For this estimate, we simplify the picture to the following: we assume
that all the impurities whose distance to the hole localization center
is smaller than $R_{\textrm{p}}$ have their spins set by the hole
spin, and that all the impurities beyond $R_{\textrm{p}}$ belong to
the random continuum. In this ``setup'' the estimation for
$\textbf{B}_{\textrm{rand}}$ is straightforward and yields:
\begin{subequations}
\label{Brand-both}
\begin{eqnarray}
\label{Brand1}
B^{(1)}_{\textrm{rand}}\sim \sqrt{n_{\textrm{i}} L R_{\textrm{p}}^2}
J_0\exp\left(-\frac{R_{\textrm{p}}}{L}\right)\\
\label{Brand2}
B^{(2)}_{\textrm{rand}}\sim n_{\textrm{i}}^{1/3}  R_{\textrm{p}}
J^{\textrm{AF}}_0\exp\left(-\frac{1}{n_{\textrm{i}}^{1/3}a}\right)
\end{eqnarray}
\end{subequations}
In these equations, the prefactors should probably be disregarded
since they are likely to be artifacts of the simplification made
above. The exponent $J_0\exp(-R_{\textrm{p}}/L)\equiv
J^{\textrm{AF}}_0\exp(-1/n_{\textrm{i}}^{1/3}a)$, however, which is
notably the same for both $B^{(1)}_{\textrm{rand}}$ and
$B^{(2)}_{\textrm{rand}}$, is highly likely correct, so it will be
taken as our estimate for $B_{\textrm{rand}}$. Since the method
employed above was rather crude, we may not be sure that a numerical
factor of the order of unity [similar to 0.86 in Eq.~(\ref{Tc})] was
not missed, so we write the equation for $B_{\textrm{rand}}$ in the
following form:
\begin{equation}
\label{Brand}
\ln\left( \frac{B_{\textrm{rand}}}{J^{\textrm{AF}}_0}\right)\sim 
-\frac{1}{n_{\textrm{i}}^{1/3}a}
\end{equation}

\subsection{Interaction of two polarons in random continuum}

Now, having considered interaction of one bound magnetic polaron with
the random medium made by interacting magnetic impurities, we turn to
interaction between two polarons surrounded by interacting magnetic
impurities and separated by distance $r$ from each other. In
Ref.~\onlinecite{KaminskiDasSarma2002} and Sec.~\ref{sec:Tcsupp} of
this paper we evaluated the interaction strength between two bound
magnetic polarons with high precision, that is up to a numerical
prefactor. Such a precision would be excessive in the case considered
in this section, since this interaction is to be compared with the
interaction with the random medium surrounding polarons, for which
only estimate (\ref{Brand}) is available. Therefore we may limit
ourselves to an estimate of the polaron interaction strength. The
(ferromagnetic) interaction comes from the interaction of each
polaron's hole with the magnetic impurities polarized by the other,
with the concentration of the latter given by Eq.~(\ref{nf}). The
strength of this interaction reads:
\begin{equation}
\label{Epolrand}
{\cal E}_{\rm eff}\left(r, \cos\theta, T \right)
\sim -
  J_0  \exp{\left( -{r \over L_{\rm loc}} \right)} \cos{\theta},
\end{equation}
with all dimensionless prefactors omitted due to the reasons outlined
above. Comparing Eq.~(\ref{Epolrand}) with Eq.~(\ref{Brand}) we
immediately come to a (natural) conclusion that two polarons at zero
temperature will have their spins aligned in the same direction if the
distance between them satisfies the condition
\begin{equation}
\label{rcrit}
r < 2 R_{\textrm{p}}\;,
\end{equation}
where $R_{\textrm{p}}$ is the zero-temperature polaron radius given by
Eq.~(\ref{Rp}), and will have their spins pointing into two
uncorrelated directions, which are determined by their interaction
with the random medium of magnetic impurities if condition~(\ref{rcrit})
does not hold.

Employing notions of the percolation theory we used above, we can
use parameter
\begin{equation}
\label{theparameter}
\mathcal{P} = \frac{n_{\textrm{h}}^{1/3}L}{n_{\textrm{i}}^{1/3}a}\;.
\end{equation}
to characterize the ground state of the system.
If $\mathcal{P}\gg 1$, the bound magnetic polarons form a
ferromagnetically-ordered  infinite cluster, and the ground state of
the system is ferromagnetic. If $\mathcal{P}\ll 1$, the interaction
with the randomly-polarized medium of magnetic impurities breaks
correlation between the cluster spins, and no long-range
ordering of bound magnetic polarons is possible. The ground state of
the system will be randomly ordered,
with occasional isolated polarized regions around bound magnetic
polarons and their finite-size clusters. Note that this statement is
in agreement with result (\ref{Tciter}) of Sec.~\ref{sec:Tcsupp},
which states that the ferromagnetic critical temperature becomes strongly
suppressed (as compared to its value when $J^{\textrm{AF}}_0=0$) exactly
when $\mathcal{P}\approx 1$. We deliberately refrain from calling this
random state a spin glass state since the words ``spin glass'' imply
some very specific properties, which this randomly ordered state may
not possess.
 
\section{Conclusion}\label{sec:concl}

\begin{figure}
\includegraphics[width=2in]{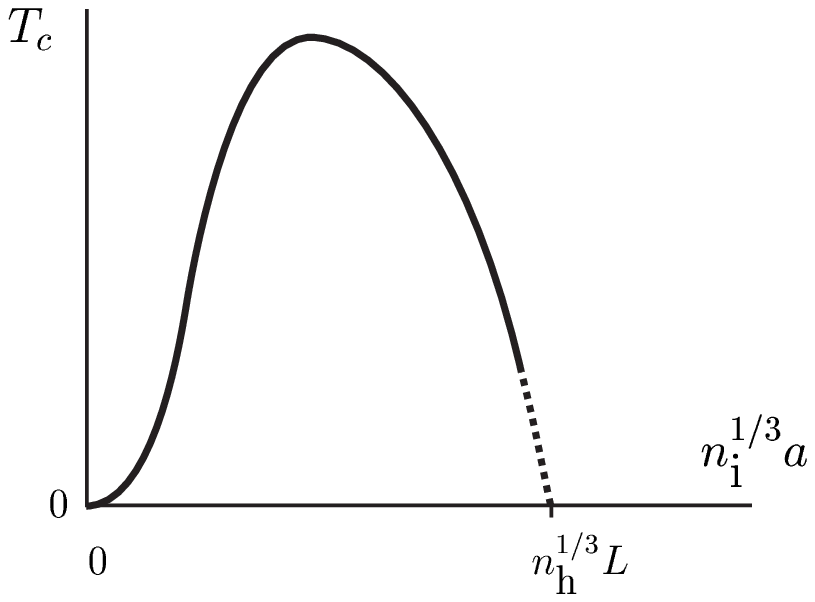}
\caption{\label{fig:phasediag} 
  Temperature of ferromagnetic transition $T_c$ as a function of
  dimensionless impurity concentration $n_{\textrm{i}}^{1/3}a$, with
  $n_{\textrm{h}}^{1/3}L = \text{const}$.  The solid line is
  determined by Eq.~(\protect\ref{Tciter}). The zero-temperature
  transition point is found in Sec.~\protect\ref{sec:zero-temp}.  The
  exact behavior of the dotted curve are beyond the scope of this
  paper.  }
\end{figure}

The results of this paper presented in Sec.~\ref{sec:Tcsupp} and
\ref{sec:zero-temp} can be summarized in Fig.~\ref{fig:phasediag}. At
very low concentration of magnetic impurities, the ferromagnetic
transition temperature is low, because of the obvious reason that the
impurity atoms mediate ferromagnetic interaction between bound
magnetic polarons. Low concentration of impurities means weak
interaction between polarons and, therefore, low transition
temperature. As the concentration of impurities goes up, the
ferromagnetic interaction between bound magnetic polarons becomes
stronger, and $T_c$ increases. At lower concentration of magnetic
impurities, antiferromagnetic interaction between them can be
neglected, due to its small decay length $a$ [Eq.~(\ref{JAF})], as
compared to the large decay length $L$ of the impurity-hole
interaction. At a certain value of the concentration of magnetic
impurities, this antiferromagnetic interaction becomes important and
eventually the interaction between the polarons weakens, and $T_c$
goes down as $n_{\textrm{i}}^{1/3}a$ grows, see Eq.~(\ref{Tciter}).
We note that such a non-monotonic dependence of the ferromagnetic
transition temperature was observed in experiment [see, {\em e.~g.}
Ref.  \onlinecite{Tcexp}]. At a certain value of the concentration of
magnetic impurities the Curie temperature is suppressed to zero and a
quantum phase transition from a ferromagnetic state into a valence
bond glass state is anticipated [see Sec.~\ref{sec:zero-temp}]. The
actual nature of this disordered valence bond phase remains an
interesting topic for future theoretical studies.

We have also argued in this paper that, although the DMS ferromagnetic
transition is strictly speaking \emph{not} a percolation transition,
the polaron percolation theory should provide a reasonable estimate
for the ferromagnetic transition temperature $T_c$, see
Sec.~\ref{sec:perc}.

\begin{acknowledgements}
This work is supported by US-ONR and DARPA.
\end{acknowledgements}

\bibliography{qpt}

\end{document}